# Virtual Identification of Essential Proteins Within the Protein Interaction Network of Yeast

## Ernesto Estrada

*Complex Systems Research Group*, X-Ray Unit, RIAIDT, University of Santiago de Compostela, Edificio CACTUS, Campus Sur, Santiago de Compostela 15782, Spain.

Tel. +34 981 563 100 Ext. 16262
Fax: +34 981 547 077
E-mail: estrada66ahoo.com

Topological analysis of large scale protein-protein interaction networks (PINs) is important for understanding the organisational and functional principles of individual proteins. The number of interactions that a protein has in a PIN has been observed to be correlated with its indispensability. Essential proteins generally have more interactions than the non-essential ones. We show here that the lethality associated with removal of a protein from the yeast proteome correlates with different centrality measures of the nodes in the PIN, such as the closeness of a protein to many other proteins, or the number of pairs of proteins which need a specific protein as an intermediary in their communications, or the participation of a protein in different protein clusters in the PIN. These measures are significantly better than random selection in identifying essential proteins in a PIN. Centrality measures based on graph spectral properties of the network, in particular the subgraph centrality, show the best performance in identifying essential proteins in the yeast PIN. Subgraph centrality gives important structural information about the role of individual proteins and permits the selection of possible targets for rational drug discovery through the identification of essential proteins in the PIN.

# 1    Introduction

A recent explosion of research papers related to the structure of complex networks has led to important results related to the topological properties of biological networks [1]. These networks, which include metabolic networks [2, 3] and protein interaction networks (PINs) [4-6], share important structural features with other real-world networks in disparate fields ranging from the Internet to social networks [7-9]. One of the most important topological characteristics shared by almost all complex networks is the so-called "small-worldness" [10]. Other properties, such as "scale-freeness" [11] and modularity [12, 13] have also been reported for metabolic networks and PINs. On the other hand, the number of links per node (node degree) in the PIN of *S. cereviciae* has been observed to be correlated with the lethality of removing such proteins from the PIN [14]. According to this result those nodes with a large number of links, the so-called *hubs*, tend to be essential. The indispensability of a gene defines its functional significance and when such genes are knocked out the cell becomes unviable [15]. The yeast PIN has also been the objective of several other topological analyses aimed at the detection of protein functionality and evolution [16-19].

Node degree is one of the several local topological properties of networks that are known as *centrality measures*. The notion of centrality comes from its use in social networks [20]. Intuitively, it is related to the ability of a node to communicate directly with other nodes, or to its closeness to many other nodes or to the quantity of pairs of nodes which need a specific node as intermediary in their communications [21]. These ideas have materialised in some well-known centrality measures such as degree centrality (*DC*), closeness centrality (*CC*), and betweenness centrality (*BC*) [20, 22]. Other centrality measures, known as eigenvector centrality (*EC*) and information centrality (*IC*), were developed by Bonacich [23, 24] and Stephenson and Zelen [25], respectively. More recently Estrada and Rodríguez-Velázquez [26] introduced a centrality measure that accounts for the weighted participation of nodes in all subgraphs of the network.

Although some centrality measures have been analysed for biological networks [27, 28] a systematic study of the relationships between centralities and protein indispensability in PINs has not been reported. It has been argued that "the biological consequences of these topological properties are not clear—in fact, there might not be any, as both small-world and scale-free behavior can be explained by well-known evolutionary events" [5]. However, we consider here a more pragmatic approach. The main objective of this work lies in analysis of the efficacy of different approaches based on centrality measures to identify lethal proteins in a protein-protein interaction network. If we have determined the PIN of a pathogenic organism for which we are interested in designing a new drug our next objective will be to select some proteins from the organism that can be used as targets for this new drug to be designed. In other words, we need to select some essential proteins that can be attacked by the new drug thus killing the pathogenic organism. In order to know which centrality measure is more appropriate for selecting essential proteins in a PIN we study here the relationships between *DC*, *CC*, *BC*, *EC*, *IC* and *SC* with essentiality of proteins in the yeast PIN.

## 2    Materials and methods
### 2.1 Centrality measures

In the context of protein-protein interaction networks (PINs) we will refer to *protein centrality* to characterize the importance or contribution of an individual protein to the global structure or configuration of the PIN.

The simplest of all centrality measures is the *degree centrality* (DC). $DC(i)$ is the number of ties incident upon a node $i$, i.e., the number of proteins that are interacting with protein $i$. Another centrality measure is the *closeness centrality* (CC) of a connected network. The closeness centrality of protein $i$ is the sum of graph-theoretic distances from all other proteins in the PIN, where the distance $d(i, j)$ from one protein $i$ to another $j$ is defined as the number of links in the shortest path from one to the other. The closeness centrality of protein $i$ in a PIN is given by the following expression:

$$CC(i) = \frac{N-1}{\sum_j d(i, j)} \qquad (1)$$

Another popular centrality measure is *betweenness centrality* (*BC*) [29]. Betweenness centrality characterizes the degree of influence a protein has in "communicating" between protein pairs and is defined as the fraction of shortest paths going through a given node. If $\rho(i, j)$ is the number of shortest paths from protein $i$ to protein $j$, and $\rho(i, k, j)$ is the number of these shortest paths that pass through protein $k$ in the PIN, then the betweenness centrality of node $k$ is given by:

$$BC(k) = \sum_i \sum_j \frac{\rho(i, k, j)}{\rho(i, j)}, \quad i \neq j \neq k \qquad (2)$$

The *eigenvector centrality* (*EC*) introduced by Bonacich [23, 24] is defined as the principal eigenvector of the adjacency matrix $\mathbf{A}$ defining the network. It simulates a mechanism in which each node affects all of its neighbors simultaneously. The defining equation of an eigenvector is $\lambda \mathbf{e} = \mathbf{A}\mathbf{e}$, where $\mathbf{A}$ is the adjacency matrix of the graph, $\lambda$ is an eigenvalue and $\mathbf{e}$ is the eigenvector. Thus, *EC* of protein $i$ is defined as the $i$th component of the eigenvector $\mathbf{e}_1$, $e_1(i)$, that corresponds to the largest eigenvalue of $\mathbf{A}$, $\lambda_1$ (principal eigenvalue or index):

$$EC(i) = e_1(i) \qquad (3)$$

Accordingly, a protein is considered central if it has a high eigenvector score, which means that it is adjacent to other proteins that themselves have high scores.

Another structural measure of centrality in a network was introduced by Stephenson and Zelen [25] and is known as *information centrality* (*IC*). It is based on the information that can be transmitted between any two points in a connected network. If $\mathbf{A}$ is the adjacency matrix of a network, $\mathbf{D}$ a diagonal matrix of the degree of each node and $\mathbf{J}$ a matrix with all its elements equal to one, then *IC* is defined by inverting the matrix $\mathbf{B}$ defined as $\mathbf{B} = \mathbf{D} - \mathbf{A} + \mathbf{J}$ in order to obtain the matrix $\mathbf{C} = (c_{ij}) = \mathbf{B}^{-1}$ from which the information matrix is obtained as follows:

$$\mathbf{I}_{ij} = (c_{ii} + c_{jj} - c_{ij})^{-1} \qquad (4)$$

The information centrality of the protein $i$ is then defined by using the harmonic average:

$$IC(i) = \left[\frac{1}{N} \sum_j \frac{1}{\mathbf{I}_{ij}}\right]^{-1} \qquad (5)$$

Stephenson and Zelen [25] proposed to define $\mathbf{I}_{ii}$ as infinite for computational purposes, which makes $\frac{1}{\mathbf{I}_{ii}} = 0$. Newman [30] has recognized that *IC* is another closeness measure, which in essence measures the harmonic mean lengths of paths ending at a node $i$, which is smaller if $i$ has many short paths connecting it to other nodes.

Finally, we will consider a centrality measure recently introduced to account for the participation of a node in all subgraphs of the network. The *subgraph centrality* (*SC*) is defined [26] as:

$$SC(i) = \sum_{l=0}^{\infty} \frac{\mu_l(i)}{l!} = \sum_{j=1}^{N} [v_j(i)]^2 e^{\lambda_j} \qquad (6)$$

where $\mu_l(i)$ is the number of walks starting and ending at node $i$, i.e., closed walks of length l starting at $i$, $(v_1, v_2,...,v_n)$ is an orthonormal basis of $R^N$ composed by eigenvectors of $\mathbf{A}$ associated to the eigenvalues $\lambda_1, \lambda_2,..., \lambda_N$, and $v_j(i)$ is the $i$th component of $v_j$.

Accordingly, $SC(i)$ counts the total number of closed walks in which protein $i$ takes part in the PIN and gives more weight to closed walks of short lengths. Closed walks are related to the network subgraph [26, 31]. Thus, $SC$ accounts for the number of subgraphs in which a protein participates, giving more weights to smaller subgraphs, which have been previously identified as important structural motifs in biological networks [32-34].

This centrality measure can be decomposed so as to account for the contribution of closed walks of even and odd lengths in the network: $SC(i) = SC_{even}(i) + SC_{odd}(i)$ (see Appendix). While closed walks of even length can be trivial going back and further for nodes of acyclic subgraphs, closed walks of odd length are related to non-trivial subgraphs containing odd-membered cycles. The expressions for these two centralities are given below (see Appendix):

$$SC_{even}(i) = \sum_{j=1}^{N} [v_j(i)]^2 \cosh(\lambda_j) \qquad (7)$$

$$SC_{odd}(i) = \sum_{j=1}^{N} [v_j(i)]^2 \sinh(\lambda_j) \qquad (8)$$

## 2.2 Yeast PIN data set

The protein-protein interaction network (PIN) of yeast, *Saccharomyces cerevisiae*, used here, was compiled by Bu *et al.* [35]. The original data was obtained by von Mering *et al.* [36] by assessing a total of 80,000 interactions among 5400 proteins reported previously and assigning each interaction a confidence level. Bu *et al.* [35] focused on 11,855 interactions between 2617 proteins with high and medium confidence in order to reduce the interference of false positives. This interaction map is considered here as a network in which proteins are represented as the nodes and two nodes are linked by an edge if the corresponding two proteins can be expected with high or medium confidence of interacting. This PIN of 2617 nodes and 11,855 links is formed by a main cluster of 2224 proteins sharing 6608 interactions which is used here for the analysis and is illustrated in Figure 1.

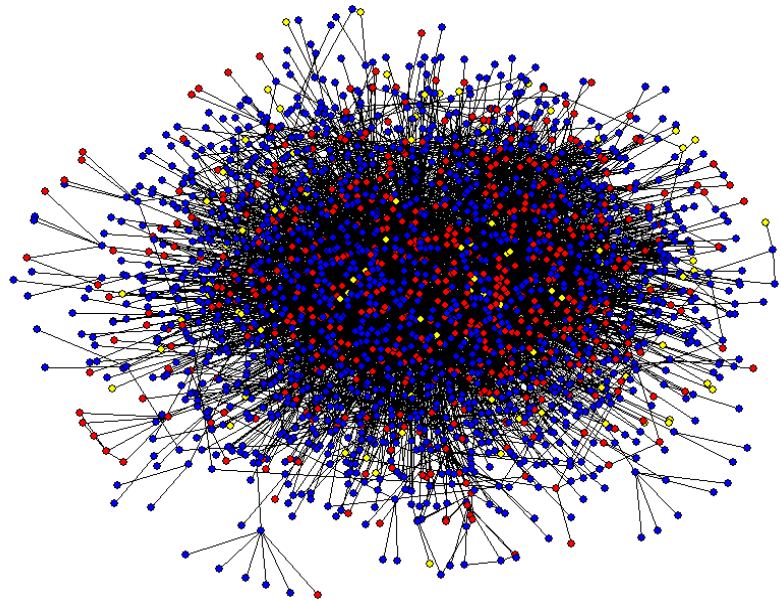

**Figure 1.** The complete PIN of yeast showing essential proteins in red and nonessential ones in blue, yellow circles correspond to proteins with unknown essentiality.

The indispensability of a protein defines the functional significance of a gene at its most basic level. Essential genes are those upon which the cell depends for its viability. Lethality can be determined without knowing the function of a gene by using, for example, random transposon mutagenesis [37] or gene-deletion [38]. Using the GENECENSUS database (http://bioinfo.mbb.yale.edu/genome/), we checked for all proteins in the main cluster of the yeast PIN for indispensability. Almost 80% of essential proteins in this main component of the yeast PIN form a connected cluster showing a high level of connectivity among essential proteins (see Figure 2).

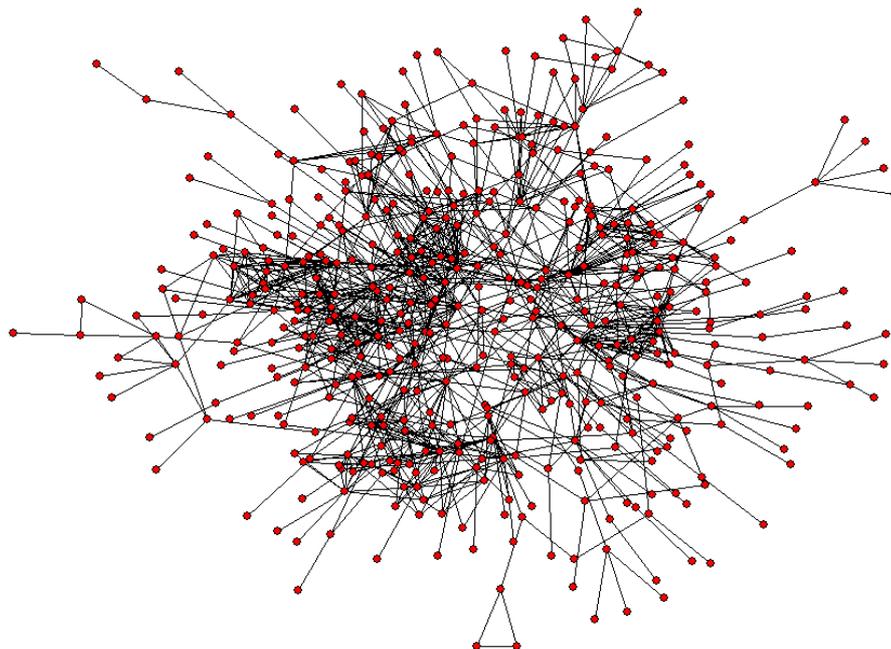

**Figure 2.** The main cluster of essential proteins of yeast containing 78.2% of all essential proteins in the PIN.

## 3      Results

The main objective of this work is to analyse the potential of centrality measures to select essential proteins in a protein-protein interaction network. Degree centrality was previously found to be correlated with protein lethality in yeast PIN in such a way that there is a larger number of essential proteins among those highly connected than among proteins with low degree [14]. Here our approach consists in ranking proteins according to their values of centrality; we select the top top 1%, top 5%, etc. of proteins, and determine how many of these are essential in the yeast PIN. For purposes of comparison we also select proteins at random taking the average of the number of essential proteins after 20 random selections. We first determined that from the three subgraph centrality measures ($SC$, $SC_{even}$ and $SC_{odd}$) the most effective detection of essential proteins is always obtained with the odd subgraph centrality, which for brevity will be denoted as $SCo$. In Figure 3 we illustrate the number of essential proteins detected by $SCo$, $DC$, $CC$, $BC$, $EC$ and $IC$ as well as by the random selection method up to the top 25% of proteins in the PIN, which represents an amount equivalent to the total number of essential proteins in the main cluster of the yeast PIN.

As can be seen in Figure 3, all centrality measures perform significantly better than the random selection method in selecting essential proteins in the yeast PIN. This result is important from the practical point of view because it means that the use of centrality measures for selecting essential proteins in a PIN is significantly more efficient than random selection. If we select these proteins at random we have a high chance of detecting a low ratio of essential proteins, and losing time, efforts and resources. However, the use of centrality measures to rank the proteins in this PIN and selecting, for instance, the top 1% of them will guaranty a higher probability of selecting essential proteins.

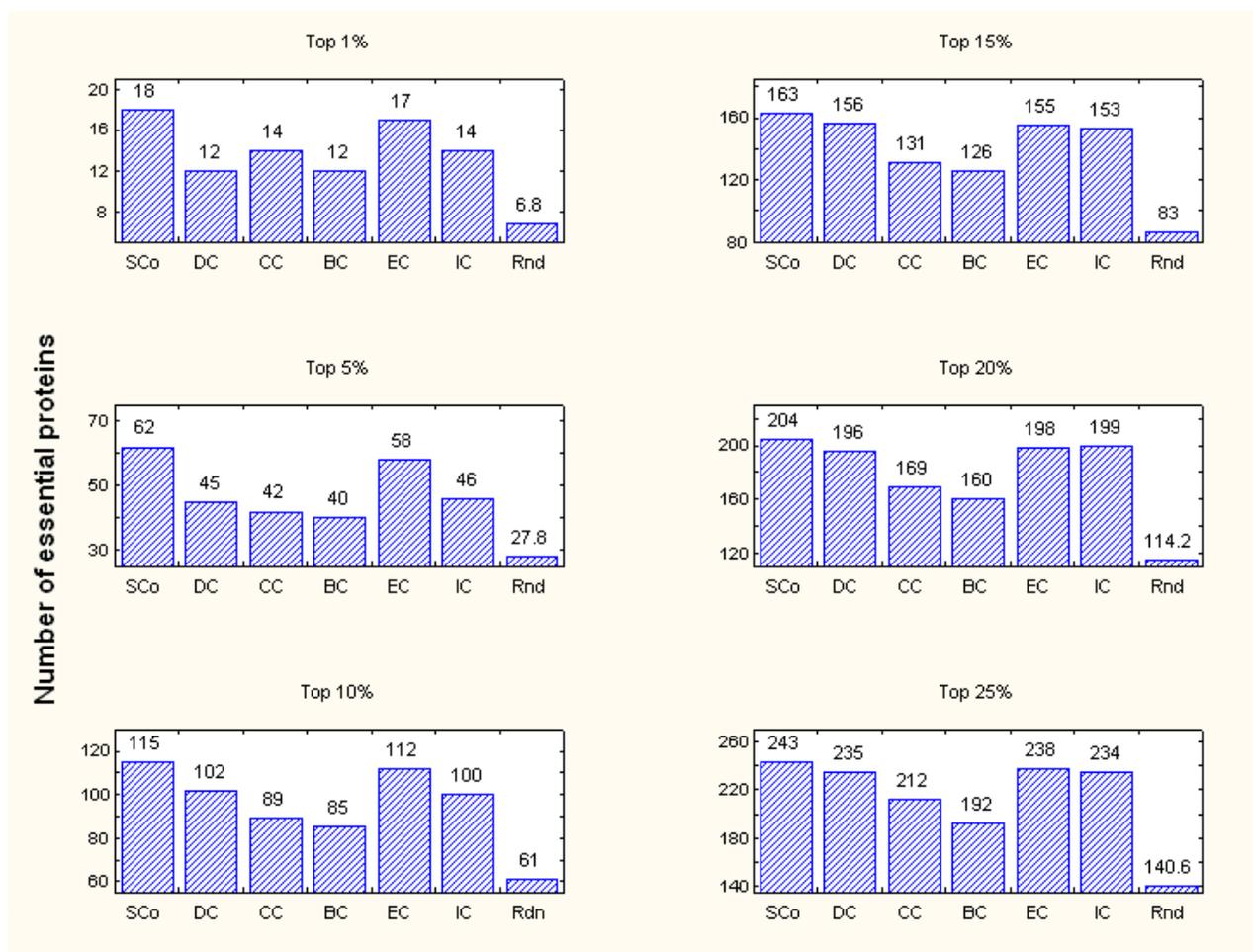

**Figure 3.** Number of essential proteins selected by ranking proteins according to their values of centrality and at random (after 20 realizations).

An individual analysis of each centrality measure reveals interesting characteristics of this selection parameter. In all cases the subgraph centrality measure *SCo* identifies the larger number of essential proteins in the PIN among all centrality measures, followed very closely by the other spectral centrality measure, *EC*. The poorest performances are obtained by using *BC* and *CC*, while *DC* and *IC* report numbers of essential proteins intermediate between those of *SCo-EC* and *BC-CC*. Any centrality measure is capable of identifying almost twice as many indispensable proteins as the random selection. If we consider the top 1% of proteins in the PIN to be comprised of 30 proteins, *SCo* is able to identify 60% of these proteins as essential, 20% more than identified by *DC*. These differences are less marked as the percentage of proteins selected increases. For instance, when the top 25% of proteins are selected, *SC* identifies 8 essential proteins more than *DC* and only 5 more than *EC*. If we are interested in selecting a few proteins to test their lethality so as to select some of them as targets for new drugs the analysis of the top 1% of proteins ranked by centrality measures will be more appropriate than analysis of the top 25%. However, the difference with respect to *BC* and CC is still significant because *SC* identifies 51 essential proteins more than *BC* and 31 more than *CC*. At this point *SCo* overtakes the random selection method for more than 100 essential proteins.

The proteins selected by each of the centrality measures studied together with the links between them can be considered as sub-networks of the yeast PIN. For instance, if we take the top 1% proteins selected by any centrality measure we can form a network of 30 nodes together with the links joining them. These networks are not necessarily connected. Suppose that the top $x$ nodes ranked by a centrality measure are topologically distant from each other with no link joining any pair of such nodes. In this case we will have a network formed by $x$ disconnected nodes. However, if a link exists between each pair of these nodes the resulting sub-network will be connected. The study of the topological characteristics of these sub-networks can shed some light on the place that essential proteins occupy in the yeast PIN. In Table 1 we give the values of three topological characteristics for these sub-networks formed by the nodes ranked as the top 1%, 10% and 25% according to the different centrality measures as well as the random selection method. These topological parameters are the Watts and Strogatz [10] clustering coefficient, $C$; the average distance among reachable pairs of nodes, $L$, i.e., the number of edges in the shortest path between each pair of nodes; and the average degree $\langle k \rangle$.

**Table 1.** Topological characteristics of the sub-network formed by the top 25% of proteins selected by centrality measures or at random.

| Method | | C | L[a] | <k> |
|---|---|---|---|---|
| SC | 1% | 0.529 | 1.586 | 12.67 |
| | 10% | 0.391 | 2.418 | 11.84 |
| | 25% | 0.217 | 3.052 | 10.44 |
| DC | 1% | 0.306 | 3.203 | 3.00 |
| | 10% | 0.253 | 2.905 | 7.49 |
| | 25% | 0.200 | 3.140 | 11.05 |
| BC | 1% | 0.032 | 3.120 | 2.47 |
| | 10% | 0.130 | 2.902 | 6.05 |
| | 25% | 0.112 | 3.310 | 9.18 |
| CC | 1% | 0.277 | 2.568 | 4.60 |
| | 10% | 0.132 | 2.498 | 8.78 |
| | 25% | 0.142 | 2.987 | 10.74 |
| EC | 1% | 0.529 | 1.586 | 12.67 |
| | 10% | 0.330 | 2.210 | 12.03 |
| | 25% | 0.210 | 3.052 | 10.27 |
| IC | 1% | 0.361 | 3.066 | 4.06 |
| | 10% | 0.243 | 2.681 | 8.93 |
| | 25% | 0.198 | 3.103 | 11.34 |
| Random | 1% | 0.000 | -[b] | 0.00 |
| | 10% | 0.000 | -[b] | 0.22 |
| | 25% | 0.146 | 5.867 | 1.42 |

[a]Average distance among reachable pairs of nodes. [b]Very low number of connected nodes

The sub-networks formed by nodes selected by the two spectral methods show larger clustering coefficients and lower average distances among all the sub-networks studied. As expected, the sub-networks formed by random selection of nodes are highly disconnected, showing very low clustering. The clustering coefficient measures the degree of cliquishness of a node in the network and the sub-networks formed by selecting nodes according to their higher *SC* and *EC* show the largest cliquishness of all the sub-networks studied. Coincidentally, these two selection methods are the most efficient for the identification of essential proteins in the yeast PIN.

## 4        Discussion

The fact that centrality measures overtake the random selection in detecting essential proteins clearly indicate that these topological measures encode important structural information related to protein lethality, i.e., they are describing some fundamental topological information of the PIN which is likely to be essential for protein function. Our results also indicates that protein indispensability does not depend on how close a protein is to many other proteins, nor on the number of pairs of proteins which a particular protein needs as intermediary in their communications in the protein-protein interactions. More importantly the protein essentiality appears to be related on how much a protein is implicated in clusters of proteins forming odd-membered cycles, such as triangles, pentagons, etc. This hypothesis is confirmed in part by the fact the proteins selected by any of the spectral measures of centrality form clusters of highly interconnected nodes showing a high number of triangles as measured by the clustering coefficient. These results agree with those reported by Yu *et al.* [15] who determined that within the interaction network, essential proteins tend to be more cliquish.

As we know, *SCo* takes into account not only the number of triangles but also the number of other odd-cyclic subgraphs in which a node participates. This can explain the superior performance of *SCo* in the selection of essential proteins compared to the other centrality measures. These subgraphs — particularly triangles — can play an important role in an understanding of the evolution of the protein–protein interaction network [32-34]. According to the coupled duplication-divergence model of evolution after gene duplication, both of the expressed proteins will have the same interactions [39]. In this model, it is proposed that both duplicate genes are subject to degenerative mutations, losing some functions but jointly retaining the full set of functions present in the ancestral gene. More recently, van Noort *et al.* [40] have reproduced the scale-free and small-world characteristics of the yeast co-expression network using a similar model, based on the simple neutralist's model, which consists of co-duplication of genes with their transcription factor binding sites (TFBSs), deletion and duplication of individual TFBSs, and gene loss [40]. Among the effects manifested by these models on the topology of the PIN is the tendency to generate bi-connected triplets and quadruples of nodes; i.e., triangles and squares. Triangles are formed among the duplicating genes and any neighbour of the parent gene, and squares are formed analogously between duplicating genes and any pair of neighbours of the parent gene. Triangles can also be involved in a great number of other subgraphs, which are all well accounted for by *SCo*. Consequently, *SCo* shows the best performance in selecting essential proteins in the yeast PIN probably because the indispensability of a given protein in the PIN is related to its implications in certain essential structural motifs containing odd-cyclic structures.

## 5        Conclusions

We have shown that the lethality associated with removal of a protein from the yeast proteome correlates with different centrality measures of the nodes in the PIN. These measures, which reflect different topological characteristics of networks, are significantly better than random selection in identifying essential proteins in a PIN. Our results confirm that the indispensability of a given protein in a PIN not only depends on the individual biochemical function and genetic redundancy of the protein but also on the global organisation of interactions and the topological role of such proteins in the network [14].

The centrality measures based on graph spectral properties of the network, in particular the subgraph centrality, show the best performance in identifying essential proteins in the yeast PIN. The odd subgraph centrality ($SCo$) quantifies the participation of a node in the different odd-cyclic subgraphs of the network, assigning more importance to the smaller ones, which have been identified as important motifs in biological networks [32-34]. Consequently, the use of $SCo$ in studying protein-protein interaction networks can provide important structural information about the role of individual proteins in the global organisation of protein interactions. At the same time ranking proteins by means of this centrality measure can be an effective method for selecting possible targets for rational drug discovery through the identification of essential proteins in a PIN.

## Appendix

The $i$th diagonal entry of the $l$th power of the adjacency of the network $A^l$ is given by:

$$\mu_l(i) = \sum_{j=1}^{N} \left(\lambda_j\right)^l \left[v_j(i)\right]^2 \tag{A1}$$

Then, by substituting (A1) in the expression defining $SC(i)$ we obtain:

$$SC(i) = \sum_{l=0}^{\infty} \left\{ \sum_{j=1}^{N} \frac{\left(\lambda_j\right)^l \left[v_j(i)\right]^2}{l!} \right\} = \sum_{j=1}^{N} \left[v_j(i)\right]^2 \sum_{k=0}^{\infty} \frac{\left(\lambda_j\right)^l}{l!} \tag{A2}$$

Now we can divide the terms in the infinite sum of expression (A2) into two groups, one grouping the "even" powers — describing closed walks of even length — and the other grouping the "odd" ones — describing closed walks of odd length —. Then we obtain the following expression:

$$SC(i) = \sum_{j=1}^{N} \left[v_j(i)\right]^2 \sum_{l=0}^{\infty} \left( \frac{\left(\lambda_j\right)^{2l}}{2l!} + \frac{\left(\lambda_j\right)^{2l+1}}{(2l+1)!} \right) = \sum_{j=1}^{N} \left(v_j^i\right)^2 \sum_{k=0}^{\infty} \frac{\left(\lambda_j\right)^{2k}}{2k!} + \sum_{j=1}^{N} \left(v_j^i\right)^2 \sum_{k=0}^{\infty} \frac{\left(\lambda_j\right)^{2k+1}}{(2k+1)!} \tag{A3}$$

Using the McLaurin series for the infinite sums of (A3) we obtain the expressions for $SC_{even}(i)$ and $SC_{odd}(i)$:

$$SC(i) = \sum_{j=1}^{N} \left[v_j(i)\right]^2 \cosh\left(\lambda_j\right) + \sum_{j=1}^{N} \left[v_j(i)\right]^2 \sinh\left(\lambda_j\right) = SC_{even}(i) + SC_{odd}(i) \tag{A4}$$

*The author thanks the "Ramón y Cajal" program, Spain for partial financial support.*